\documentclass{iaus}
\usepackage{graphicx}

\title[Homogeneous studies of transiting planets] 
{Homogeneous studies of transiting planets: an online catalogue}

\author[John Southworth]   
{John Southworth$^1$}

\affiliation{$^1$\,Astrophysics Group, Keele University, Staffordshire, ST5 5BG, UK \\ email: {\tt jkt@astro.keele.ac.uk}}

\pubyear{2012}
\volume{282}
\pagerange{1--2}
\jname{From Interacting Binaries to Exoplanets: Essential Modelling Tools}
\editors{A. Editor, B. Editor \& C. Editor, eds.}
\begin{document}

\maketitle

\begin{abstract}
The derived physical properties of the known transiting extrasolar planetary systems come from a variety of sources, and are calculated using a range of different methods so are not always directly comparable. I present a catalogue of the physical properties of 58 transiting extrasolar planet and brown dwarf systems which have been measured using homogeneous methods, resulting in quantities which are internally consistent and well-suited to detailed statistical study. The main results for each object, plus a critical compilation of literature values for all known systems, have been placed in an online catalogue. TEPCat can be found at: {\tt http://www.astro.keele.ac.uk/$\sim$jkt/tepcat/}
\end{abstract}

\firstsection

\section{The {\it Homogeneous Studies} project}

At this point roughly 130 transiting extrasolar planets (TEPs) are known, discovered by over 20 different groups and consortia. The characterisation of these objects is complicated by the fact that the number of measured quantities needed to calculate their physical properties is one greater than the number available directly through photometric and spectroscopic observations. This leads to the requirement to include additional constraints, typically from theoretical stellar models, in order to arrive at a determinate solution. The intricacy of this process leads expectedly to an inhomogeneity in the resulting solutions, and hinders statistical studies of these objects. In order to expedite this problem I am undertaking a project to measure the physical properties of the known TEPs by applying homogeneous methods to published data.

The transit light curves are modelled using the {\sc jktebop} code (Southworth et al., 2004a), which represents the star and planet as biaxial spheroids. Careful attention is paid to the calculation of robust statistical errors using Monte Carlo simulations (Southworth et al., 2004b), the assessment of systematic errors by a residual-permutation algorithm (Jenkins et al., 2002), issues related to the treatment of limb darkening, contaminating `third' light, and numerical integration over long exposure times. These aspects are discussed in detail in Papers I and III (Southworth 2008, 2010).

Determination of the physical properties of the TEPs is achieved by using each of five different sets of tabulated predictions from theoretical stellar models as constraints. The input quantities are the orbital velocity amplitude, spectroscopic temperature and metallicity of the parent star, plus the results calculated from the light curve solutions. The output parameters comprise the physical properties of the system (Paper\,II -- Southworth 2009). Statistical errors are supplied via a propagation analysis (Southworth et al., 2005) and systematic errors from consideration of the spread of results obtained using the five different stellar model predictions. Three quantities are free of systematic errors: the mean density of the star (Seager \& Mall\'en-Ornela 2003), and the surface gravity and equilibrium temperature of the planet (Southworth et al., 2007; Paper\,III).

In Paper\,IV (Southworth 2011) I have extended the number of objects to 58, which now includes 15 CoRoT systems, 10 observed by {\it Kepler}, five HAT discoveries, eight WASP objects, and all OGLE, TrES and XO planets. A plot of the masses and radii of the planets and their parent stars is shown in Fig.\,1. Paper\,IV also introduced an online catalogue of transiting planets, TEPCat, at {\tt http://www.astro.keele.ac.uk/$\sim$jkt/tepcat/}. \\ TEPCat consists of three parts: \\

\begin{itemize}
\item {\bf Part 1} is a critical compilation of the physical properties of all known (published) transiting planetary and brown-dwarf systems.
\item {\bf Part 2} brings together the main results from my {\it Homoegeneous Studies} project.
\item {\bf Part 3} summarises the basic observable quantities of all known transiting planets, including sky position, orbital ephemeris, and the depth and duration of the transits. \\
\end{itemize}

\noindent Each part of TEPCat is available in {\tt html} table (both with and without errorbars), {\tt ascii} and {\tt csv} formats. In the future I will add additional information such as the set of physical constants used in the project, and plots of transiting planet properties.

\begin{figure}[t]
\includegraphics[width=\textwidth]{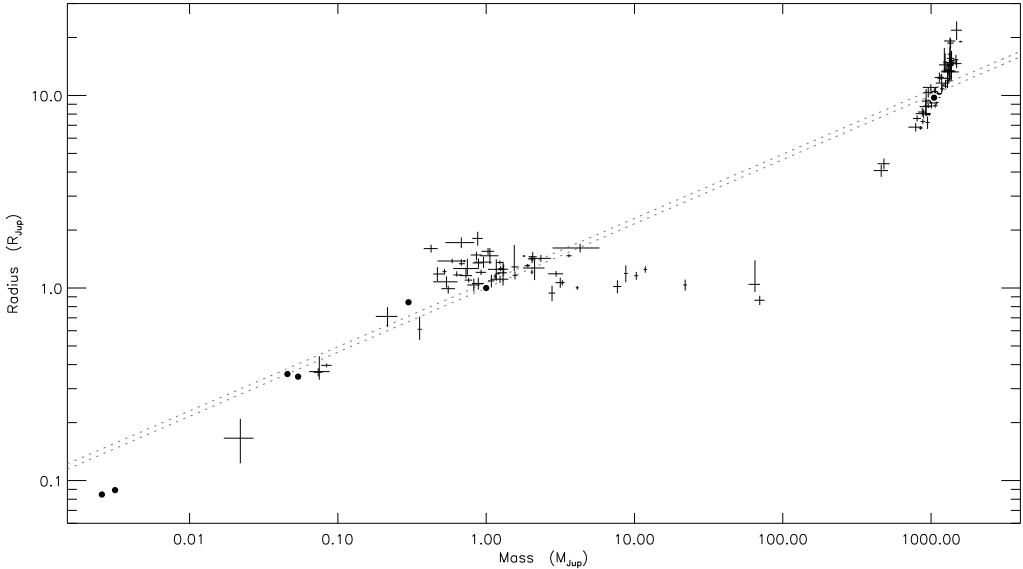}
\caption{Mass--radius plot of the 58 systems studied within this project. Several Solar system bodies are shown by filled circles. Dashed lines show the densities of Jupiter and the Sun.}
\end{figure}

\end{document}